\documentstyle[preprint,aps]{revtex}
\tightenlines
\def\journal#1, #2, #3#4, #5#6#7#8    {
    {#1~} {#2} (#5#6#7#8) #3#4}

\def\prl{\journal Phys. Rev. Lett., }

\def\npb{\journal Nucl. Phys. B, }
\def\plb{\journal Phys. Lett. B, }

\def\mpla{\journal Mod. Phys. Lett. A, }
\def\ijmpa{\journal Int. Jour. Mod. Phys. A, }
\def\jpsj{\journal J. Phys. Soc. Jnp., }
\def\jmp{\journal J. Math. Phys., }

\def\jhep{\journal J. High Energy Phys., }

\newcommand{\beq}[1]{\begin{equation}\label{#1}}
\newcommand\eeq{\end{equation}}
\newcommand{\ba}[1]{\begin{eqnarray}\label{#1}}
\newcommand{\baa}{\begin{eqnarray}}
\newcommand\ea{\end{eqnarray}}
\newcommand{\bee}{\begin{equation}}
\newcommand{\br}[1]{\overline #1}
\def\nn{\nonumber \\}

\def\n{\nu}

\newcommand{\h}{Hamiltonian}

\def\hlf{\frac{1}{2}}

\begin{document}
\title{Bosonic realization of algebras in the Calogero model}
\author{Larisa Jonke and Stjepan Meljanac
\footnote{e-mail address: 
larisa@thphys.irb.hr \\ 
\hspace*{3cm} meljanac@thphys.irb.hr}} 
\address{Theoretical Physics Division,\\
Rudjer Bo\v skovi\'c Institute, P.O. Box 180,\\
HR-10002 Zagreb, CROATIA}
\maketitle
\begin{abstract}
We study an $N$-body Calogero model in the 
$S_N$-symmetric subspace of the positive definite Fock space.
We construct a new algebra of $S_N$-symmetric operators represented
 on the symmetric
 Fock space, 
 and find a natural 
orthogonal basis by mapping the algebra onto the Heisenberg algebra.
Our main result is the bosonic realization of  nonlinear 
dynamical symmetry algebra describing the 
structure of degenerate levels of the Calogero model.

\end{abstract}
\vspace{1cm}


The Calogero model\cite{c} describes a system of $N$ bosonic particles on a
line interacting through the inverse square and harmonic potential.
It is completely
integrable in both the classical and quantum case,
the spectrum is known and the wave functions are  given implicitly.
 The model is connected with a host of physical problems,
ranging from condensed matter physics to gravity and string theory, so 
there is
considerable interest in finding the basis set of
orthonormal eigenfunctions, and the structure of the dynamical algebra
that characterizes the eigenstates of the system.

Investigations of 
the algebraic properties of the Calogero model in terms of 
the $S_N$-extended Heisenberg algebra\cite{all} defined a basic 
algebraic setup for further research.
For the Calogero model the orthonormal eigenfunctions
in terms of (Hi-)Jack polynomials\cite{st} were constructed\cite{ll}.
The
authors of Ref.\cite{2body}  showed that the dynamical symmetry algebra
of the two-body Calogero 
model was a polynomial generalization of the $SU(2)$ algebra.
The  three-body problem was also treated\cite{3body},
and the  algebra of polynomial type and the action of its generators on the
orthonornal basis were obtained.
It was  shown that in the  two-body case the polynomial $SU(2)$
algebra could be linearized, but an attempt  to generalize this result to the
$N$-body case  led to $(N-1)$ linear $SU(2)$ subalgebras that operated only on
subsets of the degenerate eigenspace\cite{ind}.
The general construction of the dynamical symmetry algebra was given in 
Ref.\cite{before}. It was shown that algebra is intrinsicaly polynomial, and 
examples for lower $N$ were provided, showing how everything could be 
computed in detail.
However it would be interesting to determine the proper elements labelling 
the energy eigenstates and find corresponding ladder operators, for 
any $N$.    

In this letter 
we define  the physical $S_N$-symmetric Fock space in terms of the
minimal number of  $S_N$-symmetric operators, and 
 find the minimal set of relations
 which define the corresponding algebra ${\cal A}_N$ of operators
$\{A_k,A_k^{\dagger}\}$.
We construct a mapping from ordinary Bose oscillators satisfying the 
Heisenberg algebra to operators $\{A_k,A_k^{\dagger}\}$, and vice versa,
for $\n>-1/N$. This mapping induces a natural orthogonal basis 
in $S_N$-symmetric Fock space, labelling the energy eigenstates in 
terms of free oscillators quantum numbers. Finally, 
we construct a symmetry algebra describing the structure of
degenerate  eigenspace, in terms of ladder operators build out 
of bosonic oscillators.

The Calogero model is defined by the following \h :
\beq 1
H=-\hlf\sum_{i=1}^N\partial_i^2+\hlf\sum_{i=1}^N x_i^2+\frac{\n(\n-1)}{2}
\sum_{i\neq j}^N\frac{1}{(x_i-x_j)^2}.\eeq
For simplicity, we have set $\hbar$, 
the  mass of particles and the frequency of harmonic 
oscillators equal to one. The dimensionless constant $\n$ is 
the coupling constant (and/or the statistical parameter) 
and $N$ is the number of particles.
The ground-state wave function is, up to normalization,
\beq 2
\psi_0(x_1,\ldots,x_N)=\theta(x_1,\ldots,x_N)\exp\left(-\hlf
\sum_{i=1}^N x_i^2\right),\eeq
where
\beq 3
\theta(x_1,\ldots,x_N)=\prod_{i<j}^N|x_i-x_j|^{\n},\eeq
with the ground-state energy $E_0=N[1+(N-1)\n]/2$.

Let us introduce the following analogs of creation and annihilation 
operators\cite{all}:
\ba 4
a_i^{\dagger}&=&\frac{1}{\sqrt{2}}\left(-\partial_i-\n\sum_{j,j\neq i}^N
\frac{1}{x_i-x_j}(1-K_{ij})+x_i\right),\nn
a_i&=&\frac{1}{\sqrt{2}}\left(\partial_i+\n\sum_{j,j\neq i}^N\frac{1}{x_i-x_j}
(1-K_{ij})+x_i\right). \ea
The elementary generators $K_{ij}$ of symmetry group $S_N$ exchange 
labels $i$ and $j$:
\ba 5
&&K_{ij}x_j=x_iK_{ij},\; K_{ij}=K_{ji},\; (K_{ij})^2=1,\nn
&& K_{ij}K_{jl}=K_{jl}K_{il}=K_{il}K_{ij},\;{\rm for}\;i\neq j,\;
i\neq l,\;j\neq l,\ea
and we set $a_i|0\rangle=0$ and $K_{ij}|0\rangle=|0\rangle$.
One can easily check that the commutators of creation and 
annihilation operators (\ref{4}) are 
\beq 6 
[a_i,a_j]=[a_i^{\dagger},a_j^{\dagger}]=0,\;
[a_i,a_j^{\dagger}]=\left(1+\n\sum_{k=1}^NK_{ik}\right)\delta_{ij}-\n K_{ij}.
\eeq
After performing a similarity transformation on the \h \ (\ref{1}), we 
obtain the reduced \h \
\beq 7
H'=\theta^{-1}H\theta=\hlf\sum_{i=1}^N\{a_i,a_i^{\dagger}\}=
\sum_{i=1}^Na_i^{\dagger}a_i+E_0 ,\eeq
acting on the space of symmetric functions. 
We restrict the Fock space $\{a_1^{\dagger n_1}\cdots a_N^{\dagger n_N}
|0\rangle\}$ 
to the $S_N$-symmetric subspace 
$F_{\rm symm}$, where ${\cal N}=\sum_{i=1}^Na_i^{\dagger}a_i$ acts as the 
total number operator. 
In the following we demand that all states have positive norm, i. e., 
$\n>-1/N$\cite{mms}.
Next, we
introduce the collective $S_N$-symmetric operators\cite{deV,before}
\beq 9 
A_n=\sum_{i=1}^N\left(a_i-\frac{B_1}{N}\right)^n\equiv\sum_{i=1}^N\br a_i^n, \;
n=2,\ldots,N .\eeq
The operator $B_1=\sum a_i$ represents
the center-of-mass operator (up to a constant) and we can remove it from the 
further considerations, since it commutes with  all $A_n^{\dagger}$  operators.
The  symmetric Fock space $F_{\rm symm}$ is $\{B_1^{\dagger n_1}
A_2^{\dagger n_2}\cdots A_N^{\dagger n_N}|0\rangle\}$ and after removing 
$B_1$ it reduces to the Fock subspace  
$\{A_2^{\dagger n_2}
\cdots A_N^{\dagger n_N}\}|0\rangle$.
The total number operator on $F_{\rm symm}$ splits into 
\ba a
&&{\cal N}=\frac{1}{N}B_1^{\dagger}B_1
+{\cal\br N},\; {\cal N}^{\dagger}={\cal N}=\sum_{i=1}^N
a_i^{\dagger}a_i,\nn
&&{\cal \br N}^{\dagger}={\cal \br N}
\equiv\sum_{k=2}^Nk{\cal N}_k .\ea
Note that ${\cal N}_k$ are the number operators of $A_k^{\dagger}$ but not of 
$A_k$, ${\cal N}_k^{\dagger}\neq {\cal N}_k$.

Let us discuss the  ${\cal A}_{N}$ algebra of the collective $S_N$-invariant
operators $A_k$ defined in Eq.(\ref{9}) and acting on $F_{\rm symm}$, 
with additions of 
$A_0=N\cdot{\bf 1}\hspace{-0.12cm}{\rm I},\;A_1=0$.  
It is obvious that 
\baa\label{Jacobi}
&&[A_m,A_n]=[A_m^{\dagger},A_n^{\dagger}]=0,\;\forall m,n, \nn
&&[A_i,[A_j,X^{\dagger}]]=[A_j,[A_i,X^{\dagger}]],
\;\forall i,j,\;{\rm any\; X^{\dagger}}.\ea
The second relation in (\ref{Jacobi}) is a consequence of the Jacobi identity.
The commutator
\beq b 
[A_m,A_n^{\dagger}]=\sum_{k=1}^{{\rm min}(m,n)}c_k(m,n)
\left(\prod \br a^{\dagger}\right)^{n-k}
\left(\prod \br a\right)^{m-k} \eeq
is an $S_N$-symmetric, normally-ordered operator, where
$c_1\sim mn$ is different from zero and does not depend on $\n$.
The symbolical expression $(\prod{\cal O})^k$ denotes a product of 
operators ${\cal O}_i$  of the total order $k$ in $\br a_i(\br a_i^{\dagger})$.
Hence, the structure of the ${\cal A}_{N}$ algebra is of the 
following type:
\beq d
[A_{i_1},[A_{i_2},\ldots,[A_{i_j},A_j^{\dagger}]]\cdots]=\sum(\prod A)^{I-j},
\eeq
where $I=\sum_{\alpha=1}^ji_{\alpha}\geq 2j$.
Generally, $j$ successive commutators of $A_{i_1},\ldots,A_{i_j}$ with 
$A_j^{\dagger}$, form  a homogeneous polynomial $\sum(\prod A)^{I-j}$ in 
$a_i$ of order $I-j$ with coefficients independent  of $\n$.
The term $A_{I-j}$ on the r.h.s. of Eq.(\ref{d}) appears with the coefficient 
$(\prod_{\alpha}i_{\alpha})j!$.
There are ${2N-1\choose N} -N$ linearly independent relations (\ref{d}).
Specially, we find
\bee\label{Jss}
[\underbrace{A_2,[A_2,\ldots[A_2}_n,A_n^{\dagger}]\ldots]=2^nn!A_n\;.\eeq
Note that any $A_n,\;n>N$ can be algebraically  expressed in terms of 
$A_m,\;m\leq N$. For example, $A_4=1/2A_2^2,\;A_5=5/6A_2A_3$ for $N\leq 3$.
The algebra ${\cal A}_N$ can be expressed in terms
of $2(N-1)$ algebraically independent operators.

For general $N$, 
the algebraic relations (\ref{d}) 
completely determine the action of $A_k, k\leq {\cal N}$  on any state:
\bee\label{nesto}
A_kA_2^{\dagger n_2}\cdots A_N^{\dagger n_N}|0\rangle=
\sum \left(\prod A^{\dagger}\right)^{{\cal N}-k}|0\rangle,\eeq
 for $k\leq \sum n_i$, reducing it to the states with $k>\sum n_i$. For 
$k>\sum n_i$, one calculates the finite set of relations from (\ref{9}) and
(\ref{6}), directly.
Hence, the ${\cal A}_N$ algebra is closed in the sense of the  successive 
commutation relation (\ref{d}), finite and of polynomial type.  
Note that the action
of $A_iA_j^{\dagger}$ (and ${\cal N}_i$)
 on the symmetric Fock subspace can be written as an 
infinite, normally ordered expansion
\bee\label{mmm}
A_iA_j^{\dagger}=\sum_{k=0}^{\infty}\left(\prod A^{\dagger}\right)^{k+j}
\left(\prod A\right)^{k+i},\; \forall i,j.\eeq
Applied to a monomial state of the finite order  ${\cal N}$ 
 in $F_{\rm symm}$, only the finite number of terms 
in Eq.(\ref{mmm}) will contribute.

The algebraic relations,  Eq.(\ref{d}), are 
independent of $\n$ and are
common to all sets of operators $\{A_k,A_k^{\dagger}\}$, with $k=2,3,\ldots,N$,
satisfying 
\bee\label{cmn}
[A_i,A_j]=[A_i^{\dagger},A_j^{\dagger}]=0,\; [{\cal N}_i,A_j^{\dagger}]=
\delta_{ij}A_j^{\dagger}.\eeq
Two different sets of operators $A(\n)$ and $A(\mu)$, satisfying the same 
algebra (\ref{d}), differ only in the generalized vacuum conditions 
for $k>\sum n_i$. Therefore, we denote the common algebra of operators 
$\{A_k,A_k^{\dagger}\}$  by ${\cal A}_N$, and its representation
for a given $\n>-1/N$ by ${\cal A}_N(\n)$. 

Let us illustrate this by the $N=3$ case. The minimal set of relations
which define the ${\cal A}_{3}(\n)$ algebra of operators $\{A_2,A_3,
A_2^{\dagger},A_3^{\dagger},{\cal \br N}\}$ is
\ba e
&&[A_i[A_2,A_2^{\dagger}]]=4iA_i,\nn
&&[A_3,[A_3,A_2^{\dagger}]]=3A_2^2,\nn
&&[A_i,[A_2,[A_3,A_3^{\dagger}]]]=6iA_iA_2,\;i=2,3,\nn
&&[A_2,[A_2,[A_2,A_3^{\dagger}]]]=48 A_3,\nn
&&[A_3,[A_3,[A_3,A_3^{\dagger}]]]=54A_3^2-\frac{9}{2}A_2^3,
\ea
plus generalized vacuum conditions
\baa\label{gvc}
&&A_2|0\rangle=A_3|0\rangle =A_2A_3^{\dagger}|0\rangle=
A_3A_2^{\dagger}|0\rangle=A_3A_2^{\dagger 2}|0\rangle=0,\nn
&&A_2A_2^{\dagger}|0\rangle =4(1+3\n)|0\rangle,\;
A_3A_3^{\dagger}|0\rangle=2(1+3\n)(2+3\n)|0\rangle,\nn
&&A_3A_2^{\dagger}A_3^{\dagger}|0\rangle=2(2+3\n)(4+3\n)A_2^{\dagger}
|0\rangle,\;
A_3A_3^{\dagger 2}|0\rangle=2(2+3\n)(11+6\n)A_3^{\dagger}|0\rangle.\ea
This algebra Eq.(\ref{e}) with the vacuum conditions (\ref{gvc}) 
has a unique representation on $F_{\rm symm}$. Using
Eqs.(\ref{e}) and (\ref{gvc}) one finds
the action of operators $A_2$ and $A_3$ on any state in the Fock space:
\ba f
&&A_2|n_2,n_3\rangle
=3{n_3\choose 2}|n_2+2,n_3-2\rangle
+4n_2(3n_3+n_2+3\n)|n_2-1,n_3\rangle,\nn
&&A_3|n_2,n_3\rangle=2\left(n_3(2+3\n)(1+3\n+3n_2)+9{n_3\choose 2}(2+3\n+n_2)
+27{n_3\choose 3}+6n_3{n_2\choose 2}\right)\times\nn
&&\times |n_2,n_3-1\rangle +
48{n_2\choose 3}|n_2-3,n_3+1\rangle
-\frac{81}{2}{n_3\choose 3}|n_2+3,n_3-3\rangle.
\ea
The ket  $|n_2,n_3\rangle$ denotes the state
$A_2^{\dagger n_2}A_3^{\dagger n_3}|0\rangle$.

The general structure of the ${\cal A}_N$ algebra can be viewed as a 
generalization of triple operator algebras\cite{trip} to $(N+1)$-tuple 
operator algebra. For $N=2$, the ${\cal A}_2$ algebra is just 
$[A_2,[A_2,A_2^{\dagger}]]=8A_2$ for a single oscillator $A_2=(a_1-a_2)^2/2$ 
describing the relative motion in the two body Calogero model.

We point out that if
the set of operators $\{A_k,A_k^{\dagger}\}$ satisfies relations
(\ref{cmn}) there exists a mapping $A_k=f_k(b_i,b_i^{\dagger})$ 
from ordinary Bose oscillators 
$\{b_i,b_i^{\dagger}\}$ to $\{A_k,A_k^{\dagger}\}$.
Namely, the relations $[A_i^{\dagger},A_j^{\dagger}]=0$ imply that the monomial
states $A_{i_1}^{\dagger}\ldots A_{i_n}^{\dagger}|0\rangle$ differing only 
in permutations of indices are all equal. The same property holds for the 
pure bosonic states. In order to construct the mapping, 
we first identify the vacua $|0\rangle_A=|0\rangle_b\equiv
|0\rangle$ and the one-particle states $A_i^{\dagger}|0\rangle
=b_i^{\dagger}|0\rangle \sqrt{\langle0|A_iA_i^{\dagger}|0\rangle}$.
Then we  normally expand operators $A_i$ as an infinite series
in terms of monomials of free bosonic oscilators
\baa\label{gener}
A_n&=&\sum_{k=0}^{\infty}\sum_{n_i,n_j'}
\left(\prod_i b_i^{\dagger n_i}\right)\left(\prod_j b_j^{n_j'}\right)
,\;\sum_i in_i=k,\;\sum_j jn_j'=k+n.\ea
The coefficients in expansion (\ref{gener}) can be calculated from matrix 
elements of scalar products in $F_{\rm symm}$\cite{jos} by applying 
expansion
to one-particle matrix elements $\langle 0|A_iA_j^{\dagger}|0\rangle$, 
then to two-particle matrix elements
$\langle 0|A_iA_jA_k^{\dagger}A_l^{\dagger}|0\rangle$, and so on.
This mapping is invertible if there are no additional relations 
(null-norm states) in $F_{\rm symm}$ when compared with the 
bosonic Fock space $F_b$. It is sufficient to demand that all 
Gram matrices of the scalar products have the same signature (structure of the 
signs of eigenvalues) in both Fock spaces. It has been found  in Ref.\cite{mms}
that the full Fock space $\{a_1^{\dagger n_1}\cdots a_N^{\dagger n_N}
|0\rangle\}$ for $\n>-1/N$
does not contain any additional null-norm 
states when compared with the bosonic Fock space $F_b$. As $F_{\rm symm}$ is
a subspace of full Fock space, it also does not contain any additional 
null-norm states, so,  $F_{\rm symm}$ is isomorphic
to  $F_b$, and the mapping $f$ is invertible for $\n>-1/N$. 

The  operators $a_i$, defined in Eq.(\ref{4}), behave like  free 
bosonic oscillators for $\n=0$ and this 
 boundary condition uniquely determines 
all coefficients in the corresponding mapping from operators $a_i$ to 
the bosonic ones and vice versa. However, in the case of $A_i$ operators, 
there is no such  boundary condition. Hence, we have a 
freedoom of fixing the boundary condition, and this corresponds to freedom of 
choosing different orthogonal basis. Here, we  choose 
a simple and natural way of fixing the boundary condition:
\baa\label{gfix}
&&A_2^{\dagger n_2}|0\rangle =\sqrt{\frac{\langle 0|A_2^{n_2}
A_2^{\dagger n_2}|0\rangle}{n_2!}}b_2^{\dagger n_2}|0\rangle ,\nn
&&A_3^{\dagger n_3}|0\rangle \sim \sum\left(\prod b_2^{\dagger i_2}
b_3^{\dagger i_3}
\right)|0\rangle , \;2i_2+3i_3=3n_3\nn
&&\vdots \nn
&&A_N^{\dagger n_N}|0\rangle \sim \sum\left(\prod b_2^{\dagger i_2}\cdots 
b_N^{\dagger i_n}
\right)|0\rangle, \;\sum_{k=2}^N ki_k=N n_N.\ea
Only after fixing the boundary condition one can determine the coefficients 
 in Eq.(\ref{gener}) in the unique way.
For the N=3 case, we present results for the first few coefficients in 
Eq.(\ref{gener}),  up to $k+n\leq 5$, for the
operators $A_2$ and $A_3$.
First, we write  general expression (\ref{gener}), up to $k+n\leq 5$:
\baa\label{b1}
A_2&=&f_2b_2+f_{22}b_2^{\dagger}b_2^2+f_{23}b_3^{\dagger}b_2b_3
+\cdots\nn
A_3&=&f_3b_3+f_{32}b_2^{\dagger}b_2b_3+\cdots ,\ea
with unknown coefficients f. Next, we apply  Eqs.(\ref{b1})
to matrix elements defined between the
states in the symmetric Fock space ($A_2^{\dagger}|0\rangle,A_3^{\dagger}|
0\rangle, A_2^{\dagger 2}|0\rangle, A_3^{\dagger 2}|0\rangle, 
A_2^{\dagger}A_3^{\dagger}|0\rangle$)
and using relations (\ref{f}), and the boundary conditions
(\ref{gfix}) we calculate unknown coefficients.
Finally, we obtain
\baa\label{fA}
A_2&=& 2\sqrt{1+3\n}\;b_2+2\left(\sqrt{2+3\n}-\sqrt{1+3\n}\right)
b_2^{\dagger}
b_2^2+2\left(\sqrt{4+3\n}-\sqrt{1+3\n}\right)b_3^{\dagger}b_2b_3,\nn
A_3&=&\sqrt{2(1+3\n)(2+3\n)}\; b_3+
\left(\sqrt{2(4+3\n)(2+3\n)}-\sqrt{2(1+3\n)(2+3\n)}\right)b_2^{\dagger}b_2b_3,
\ea
and similarly for hermitian conjugates.
From this example is clear that the mapping is real if there are no 
negative-norm states.
The inverse mapping exists for $\n>-1/3$, and all relations (\ref{gfix}), 
and 
(\ref{fA}) can 
be reversed:
\baa
b_2&=&\frac{A_2}{2\sqrt{1+3\n}}+\frac{A_2^{\dagger}A_2^2}{8(1+3\n)^{3/2}}
\left(
\sqrt{\frac{1+3\n}
{2+3\n}}-1\right)\nn &+&\frac{A_3^{\dagger}A_2A_3}{4(1+3\n)^{3/2}(2+3\n)}\left(
\sqrt{\frac{1+3\n}{4+3\n}}-1\right),\nn
b_3&=&\frac{A_3}{\sqrt{2(1+3\n)(2+3\n)}}+\frac{A_2^{\dagger}A_2A_3}
{4\sqrt{2(1+3\n)^3(2+3\n)}}
\left(\sqrt{\frac{1+3\n}{4+3\n}}-1\right).\ea
Two different
states $A_2^{\dagger n_2}\cdots A_N^{\dagger n_N}|0\rangle$
and $A_2^{\dagger n_2'}\cdots A_N^{\dagger n_N'}|0\rangle$ with the same
energy ($\sum in_i=\sum in_i'$) are
not orthogonal. For example,
\bee
\langle 0|A_3^2A_2^{\dagger 3}|0\rangle=\langle 0|A_2^3A_3^{\dagger 2}|0\rangle=
\frac{144}{N}(N-1)(N-2)(1+\n N)(2+\n N) .\eeq
However, monomial states 
$\prod b_i^{\dagger n_i}/\sqrt{n_i!}|0\rangle$ 
in $F_b$ are orthogonal, so when we express 
$b_i=f_i^{-1}(A_k,A_k^{\dagger})$, we obtain a natural orthogonal states in 
the symmetric Fock space, 
labelled by $(n_2,\ldots,n_N)$, i.e., by free 
oscillators quantum numbers. 
Degenerate, orthogonal energy eigenstates of 
level ${\cal N}$ are then defined by ${\cal N}=\sum in_i$.
For example, some orthogonal states for $N=3$ case are 
\baa\label{orthstates}
&&b_2^{\dagger}|0\rangle=\frac{1}{2\sqrt{1+3\n}}A_2^{\dagger}|0\rangle,\;
b_3^{\dagger}|0\rangle=\frac{1}{\sqrt{2(1+3\n)(2+3\n)}}A_3^{\dagger}|0\rangle
,\nn
&&b_2^{\dagger 2}|0\rangle=\frac{1}{4\sqrt{(1+3\n)(2+3\n)}}A_2^{\dagger 2}
|0\rangle\nn
&&b_2^{\dagger}b_3^{\dagger}|0\rangle=\frac{1}
{2\sqrt{2(1+3\n)(2+3\n)(4+3\n)}}A_2^{\dagger}A_3^{\dagger}|0\rangle,\nn
&&b_2^{\dagger 3}|0\rangle=\frac{1}{8\sqrt{3(1+3\n)(2+3\n)(1+\n)}}
A_2^{\dagger 3}|0\rangle ,\nn
&&b_3^{\dagger 2}|0\rangle=\alpha\left(
-12(1+\n)A_3^{\dagger 2}|0\rangle+A_2^{\dagger 3}|0\rangle\right),
\nn &&\alpha^{-1}=12\sqrt{2(1+\n)(1+3\n)(2+3\n)
[(1+\n)(2+3\n)(11+6\n)-2]}.\ea

The dynamical symmetry algebra 
${\cal C}_N(\n)$ of the Calogero model is defined as
maximal algebra commuting with the \h \ (\ref{7}). The generators of the 
${\cal C}_N(\n)$ algebra act among the degenerate states with fixed 
energy $E={\cal N}
+E_0,$
${\cal N}$ a non-negative integer.
Starting from any of degenerate states with energy $E$,
all other states can be reached by applying generators  of the
algebra.
Degeneracy appears for ${\cal N}\geq 2$. The vacuum $|0\rangle$ and the
first excited state $B_1^{\dagger}|0\rangle$ are nondegenerate.
For ${\cal N}=2$,
the degenerate states are $B_1^{\dagger 2}|0\rangle$ and $A_2^{\dagger}
|0\rangle$; for ${\cal N}=3$, the degenerate states are
$B_1^{\dagger 3}|0\rangle$, $B_1^{\dagger}A_2^{\dagger}|0\rangle$ and
$A_3^{\dagger}|0\rangle$, etc. The number of degenerate states of level
${\cal N}$
is given by partitions ${\cal N}_1,\ldots,{\cal N}_k$ of ${\cal N}$ such that
${\cal N}=\sum_k k{\cal N}_k$.
The generators of the algebra
${\cal C}_N(\n)$ can be chosen in different ways, and in the following
we present a new and very simple bosonic realization of the 
dynamical algebra.
The generators $\{{\cal J}_i^{\pm},{\cal J}_i^0\},\;i=2,\ldots,N$ 
of the algebra 
represent a generalization of the linear 
$SU(2)$ case discussed in Ref.\cite{before}: 
\baa\label{Jg}
{\cal J}_i^+&=&\frac{1}{\sqrt{i({\cal N}_1-1)\cdots({\cal N}_1-i+1)}}
b_1^{\dagger i}b_i, \nn
{\cal J}_i^-&=&b_i^{\dagger}b_1^i\frac{1}{\sqrt{i({\cal N}_1-1)\cdots
({\cal N}_1-i+1)}}=\left({\cal J}_i^+\right)^{\dagger},\nn
{\cal J}_i^0&=&\hlf\left(\frac{{\cal N}_1}{i}-b_i^{\dagger}b_i\right).\ea
One can express the generators ${\cal J}_i^{\pm,0}$ in terms 
 of $\{A_k,A_k^{\dagger}\}$ 
using second relation in (\ref{gener}), for all $\n>-1/N$. 
The coefficients in expansion depend on the statistical parameter $\n$.
The generators $\{{\cal J}_i^{\pm},{\cal J}_i^0\}$
 satisfy the following new algebra for every $i,j=2,\ldots,N$:
\baa\label{Jalg}
\left[{\cal J}_i^0,{\cal J}_j^{\pm}\right]&=&\pm\hlf\left(\frac{j}{i}
+\delta_{ij}\right){\cal J}_j^{\pm},\nn
{\cal J}_i^+{\cal J}_j^-&-&\left(\frac{\sqrt{{\cal N}_1({\cal N}_1-i+j)}}
{ {\cal N}_1+j}\right){\cal J}_j^-{\cal J}_i^+=\delta_{ij}\left(
\frac{{\cal N}_1}{i}\right),\nn
{\cal J}_i^+{\cal J}_j^+&-&\sqrt{\frac{{\cal N}_1-i}{{\cal N}_1-j}}
{\cal J}_j^+{\cal J}_i^+=0, \;
\left[{\cal J}_i^0,{\cal J}_j^0\right]=0,\ea
Note that the generators 
${\cal J}_i^+$ and ${\cal J}_i^-$ are hermitian conjugate 
to each other, and that relations (\ref{Jalg}) do not depend on $\n$.
The ${\cal C}_N$ algebra contains $(N-1)$ linear $SU(2)$ subalgebras, i.e.,
$[{\cal J}_i^+,{\cal J}_i^-]=2{\cal J}_i^0$, $[{\cal J}_i^0,{\cal J}_i^{\pm}]=
\pm{\cal J}_i^{\pm}$.
Different $SU(2)$ subalgebras are connected in a nonlinear way, 
namely, commutation relations (\ref{Jalg}) 
have nonlinear (algebraic in ${\cal N}_1$)
deformations.

The states with  fixed energy $E=
{\cal N}$ are given  by $\prod_{i=1}^Nb_i^{\dagger n_i}/\sqrt{n_i!}|0\rangle$
with $\sum_{i=1}^Nin_i={\cal N}$. These states are orthonormal and build an 
irreducible representation (IRREP) of the symmetry algebra ${\cal C}_N$. 

The dimension of the IRREP on the states with fixed energy ${\cal N}$
 of the algebra ${\cal C}_N$ can be 
obtained recursively in $N$:
$$D({\cal N},N)=\sum_{i=0}^{[{\cal N}/N]}
D({\cal N}-Ni,N-1).$$ For example, for $N=1$, 
$D({\cal N},1)=1$, i.e., for a single oscillator, there is no degeneracy;
for $N=2$, $D({\cal N},2)=\left[\frac{{\cal N}}{2}\right]+1$, ${\cal N}\geq 0$;
for $N=3$,
$$D({\cal N},3)=\sum_{i=0}^{\left[\frac{{\cal N}}{3}\right]}\left[\frac{{\cal N}
-3i}{2}\right]+\left[\frac{{\cal N}}{3}\right]+1. $$
More specifically, for ${\cal N}=6n+i$, for some integer $n$ and $i=0,\ldots,5$:
\baa
D(6n+i,3)&=&(3n+i)(n+1),\; {\rm for}\;i=1,\ldots,5,\nn
D(6n,3)&=&3n(n+1)+1.\nonumber\ea
The structure of the IRREP's
of dynamical symmetry for $N=4$ is
\bee\begin{array}{ccccc}
b_1^{\dagger 4}|0\rangle & b_1^{\dagger 2}b_2^{\dagger}|0\rangle
& b_1^{\dagger}b_3^{\dagger}|0\rangle
& b_2^{\dagger 2}|0\rangle & b_4^{\dagger}|0\rangle \\
 & b_1^{\dagger 3}|0\rangle & b_1^{\dagger}b_2^{\dagger}|0\rangle
& b_3^{\dagger}|0\rangle &
\\ & & \hspace{-1.5cm} b_1^{\dagger 2}|0\rangle & \hspace{-1.0cm}
b_2^{\dagger}|0\rangle &  \\
 & & b_1^{\dagger}|0\rangle & &\\
 & & |0\rangle & &
\end{array}\eeq

Note that the dynamical symmetry  algebra of the 
\h \ $H_1=
\sum_{k=1}^N b_k^{\dagger}b_k$ is $SU(N)$. The corresponding generators
are $b_i^{\dagger}b_j$, and $[H_1,b_i^{\dagger}b_j]=0$ for all $i$ and $j$.
The degenerate states  are $\prod_{i=1}^Nb_i^{\dagger n_i}/\sqrt{n_i!}|0\rangle$
 with $\sum_in_i={\cal N}$.
They form a totally symmetric IRREP of $SU(N)$ described by the Young diagram
with ${\cal N}$-boxes 
$\underbrace{\fbox{\rule{0mm}{0.4mm}}\framebox[4mm]{...}\fbox{\rule{0mm}{0.4
mm}}}_{\cal N}$ \cite{Sn}. 

The states of the same Fock space $\{\prod_i b_i^{\dagger n_i}/\sqrt{n_i!}
|0\rangle\}$ are differently arranged in representations of $SU(N)$ and 
${\cal C}_N$ algebra. In respect to  $H_1$,
 degenerate states are organized into the $SU(N)$ 
$\underbrace{\fbox{\rule{0mm}{0.4mm}}\framebox[4mm]{...}\fbox{\rule{0mm}{0.4
mm}}}_{\cal N}$ symmetric  IRREP's. 
However, in respect to  $H_2=H'-E_0=\sum_{i=1}^Ni{\cal N}_i
=\sum_{i=1}^Nib_i^{\dagger}b_i$, with ${\cal N}_i\neq b_i^{\dagger}b_i$,
 degenerate states build the 
${\cal N}$-IRREP of the ${\cal C}_N$ algebra. 
In this sense, one can (formally)
say that the nonlinear  dynamical algebra ${\cal C}_N$ of the \h \ $H_2$
is related to the  boson 
realization of $SU(N)$ algebra describing the dynamical 
symmetry of the \h \ $H_1$.

We have studied the Calogero model in the harmonic potential for $N$ identical 
bosons. We have started with the Fock space of states with positive definite 
norms, and restricted ourselves to the subspace of all symmetric states.
After separating the center-of-mass coordinate, we have concentrated on the 
space $\{A_2^{\dagger n_2}
\cdots A_N^{\dagger n_N}\}|0\rangle$ of operators $\{A_k,A_k^{\dagger}\}$. 
There exist  number operators 
${\cal N}_k$ that are not hermitian, since the monomial states  in 
Fock space $F_{\rm symm}$
are not mutually orthogonal. The action of the operators
$A_k$ on the states in Fock space can be calculated using successive
commutators defining the algebra ${\cal A}_N$, and generalized 
vacuum conditions.
We have shown
that there exists a mapping from ordinary Bose oscillators to operators 
$\{A_k,A_k^{\dagger}\}$, and its inverse, provided that $\n>-1/N$.
In that way we obtained a natural 
orthogonal basis for $S_N-$symmetric Fock space, an alternative to 
Jack polynomials. 
Finally, we constructed a dynamical 
symmetry algebra describing the structure of 
degenerate eigenspace, in terms of ladder operators build out of the 
bosonic oscillators. This bosonic realization of the symmetry algebra provides 
more insight on the nature of the  degenerate energy eigenstates in the 
Calogero model.
These results can also  be applied to the finite Chern-Simons
matrix model\cite{pj}, since the observables, i.e., the  $SU(N)$ 
invariant operators
${\rm Tr}A^n$ defined in 
that model satisfy the same ${\cal A}_N$ algebra.

Acknowledgment

We would like to thank V. Bardek, M. Milekovi\'c and D. Svrtan for useful
discussions.
This work was supported by the Ministry of Science and Technology of the
Republic of Croatia under contract No. 00980103.

\end{document}